\documentclass[prl,twocolumn,epsfig,superscriptaddress,showpacs]{revtex4}
\usepackage{graphicx}

\begin{document}

\title{Spectrum of spontaneous emission into the mode of a cavity QED system}
\author{M. L. Terraciano}
\affiliation{Dept. of Physics, University of Maryland, College
Park, MD 20742-4111, U.S.A.}
\author{R. Olson}
\affiliation{Dept. of Physics, University of Maryland, College
Park, MD 20742-4111, U.S.A.}
\author{D. L. Freimund}
\affiliation{Dept. of Physics, University of Maryland, College
Park, MD 20742-4111, U.S.A.}
\author{L. A. Orozco}
\affiliation{Dept. of Physics, University of Maryland, College
Park, MD 20742-4111, U.S.A.}
\author{P. R. Rice}
\affiliation{Dept. of Physics, Miami University, Oxford, OH 45056,
U.S.A.}

\date{\today}

\begin{abstract}
We study the probe spectrum of light generated by spontaneous
emission into the mode of a cavity QED system. The probe spectrum
has a maximum on-resonance when the number of inverted atoms for
an input drive is maximal.  For a larger number of atoms $N$, the
maximum splits and develops into a doublet, but its frequencies
are different from those of the so-called vacuum Rabi splitting.

\end{abstract}


\maketitle
Spontaneous emission in cavity Quantum Electrodynamics (QED) has
generated considerable interest since the birth of the field
\cite{sanchez83}. The interaction between a single atom or $N$
atoms and a single cavity mode is different from that in free
space. The study of this interaction has led to ground-breaking
experiments in nonlinear optics, squeezing, nonclassical
correlations, and quantum information \cite{berman94}. Spontaneous
emission in cavity QED has been regarded as a dissipative process
from which information is lost at a rate ($\gamma$) to modes other
than the preferred cavity mode. Most work in cavity QED
spontaneous emission has focused on the enhancement or suppression
of the decay rate $\gamma$. An atom couples to the mode defined by
the cavity mirrors through an allowed transition at a rate $g$.
Photons escape the cavity due to imperfect mirror reflectivities
at a rate $2\kappa$. In the bad cavity limit $(\kappa>>\gamma,g)$
the resonant spontaneous emission changes from its free space
value as $\gamma\rightarrow\gamma(1+2C_1)$, with the single atom
cooperativity $C_1=g^2/(\kappa\gamma)$ \cite{carmichael91}. The
enhancement factor is related to the ratio of the atomic cross
section to the cavity mode cross section multiplied by the average
number of reflections inside the cavity. This effect broadens the
spectrum, but causes no splitting \cite{heinzen87}. There are many
experimental demonstrations of enhanced and suppressed spontaneous
emission in this regime (see for example the article by Hinds in
Ref.~\cite{berman94}). If the reflectivity of the mirrors is high
enough and the coupling between the atom and the cavity can become
comparable to the two decays ($g \approx\kappa,\gamma$),
spontaneous emission into the cavity is reversible. The total
fraction of emission out of the cavity is $(1+2C_1')$; where
$C_1'=C_1 2\kappa/(\gamma+2\kappa)$. The factor
$2\kappa/(\gamma+2\kappa)$ is the fraction of photons emitted into
the cavity mode, exiting the cavity via the mirror \cite{cui05}.

Spontaneous emission plays a dual role; it is a decohererence
source, but it is also a way to extract information out of the
system. An interrogation of the system through spontaneous
emission is an unambiguous probe of the state of the atomic part
of the atom-cavity system.  This letter presents our
investigations of the spectrum of light generated by a spontaneous
emission process into the mode a driven optical cavity. The
properties of the spontaneous emission channel in this system,
together with the cavity output,  are worth extensive study.
Cavity QED has been identified as an environment to transfer
information and entanglement between matter and light qubits
\cite{cirac97}. The information has to exit the system through one
of the two available channels as part of quantum interconnects and
information protocols.

Our cavity QED system consists of a high finesse optical resonator
where one or a few atoms interact with a single longitudinal and
transverse mode of the cavity; the resulting coupling rate $g$
depends on the dipole moment of the transition and the electric
field that carries the energy of one photon. The combination of
the rates in this system gives two dimensionless numbers, the
first measures the effect of $N$ atoms in the system through the
cooperativity $C=C_1N$ and the second measures the non-linearity
of the system through the saturation photon number
$n_0=\gamma^2/8g^2$. The non-linearity is intrinsic in the system
as a two-level atom in the excited state can not go further up the
energy ladder, but only can come down through spontaneous or
stimulated emission; in contrast, the cavity is a harmonic
oscillator and its energy can increase without bound. The system
is driven on-axis by a classical field $\varepsilon/\kappa$
normalized normalized to photon flux units:
$y=\varepsilon/(\kappa\sqrt{n_0})$. Its frequency $\omega_l$ is
detuned from the atomic resonance $\omega_a$ and cavity resonance
$\omega_c$ by an amount $\Omega=\omega_l-\omega_c$, with
$\omega_a=\omega_c$. The atomic inversion $\sigma^z$ is related to
the expectation value of the intracavity field $a$ and the
collective atomic polarization $\sigma_+$ with their Hermitian
conjugates through the equation of motion for their expectation
values:
\begin{equation}
\frac{d<\sigma^z>}{\gamma dt}=<a \sigma^+>+<a^{\dagger} \sigma^->
-(<\sigma^z>+1). \label{inversion}
\end{equation}
This equation shows that the atomic inversion is related to the
correlation between the field in the cavity and the atomic
polarization. In steady state the cross terms $<a
\sigma^+>+<a^{\dagger} \sigma^->$ are proportional to the
inversion. The probe spectrum (emission as function of driving
field frequency) is related to the steady-state magnitude of the
cross terms of Eq.~\ref{inversion} as a function of driving laser
laser frequency.

The system can be accurately modelled, for weak excitation, as
having either zero or one excitations of the coupled normal modes
of the field and the atoms. If we assume fixed atomic positions,
to first order in the excitation, $O(y^2)$, the equilibrium state
is the pure state \cite{carmichael91,brecha99}:
\begin{equation}
|\psi_{\rm ss}\rangle = |0,G\rangle + x\sqrt{n_0}|1,G\rangle
-p\sqrt{n_0}|0,E\rangle + O(y^4). \label{psiss}
\end{equation}
\noindent Here $|n,G\rangle$ represents $n$ photons with all $(N)$
atoms in their ground state, $|n,E\rangle$ represents $n$ photons
with one atom in the excited state with the rest $(N-1)$ in their
ground state. The small parameter is the expectation value of the
intracavity field in the presence of atoms; $x \sqrt{n_0}= \langle
a \rangle$. The induced atomic polarization of $N$ atoms is
$p=-2Cx$, which depends on the normalized input driving field $y$.
We can pass now from Eq.\ref{inversion} to the semiclassical
analysis of the weak field limit and use the steady state wave
function to evaluate the inversion. To lowest order we have:
$\langle a \sigma^+ \rangle=\langle a \rangle \langle \sigma^+
\rangle + O(y^4)$, which is equivalent to the decorrelation
 of the expectation values of the product
of the field and the polarization \cite{carmichael93book}, and we
recover the Maxwell Bloch equations \cite{lugiato84}.

The spectrum of the transmitted light is given by the frequency
dependent coefficients of the single excitation coefficients of
the steady state (Eq. \ref{psiss}), those of order $y$. We can use
the state equation of cavity QED to find the transmitted spectrum
in both the field and the inversion.

The optical bistability literature \cite{lugiato84} gives the
relation between the expectation values of the field and
polarization $<a>,~<\sigma^+>$ and measurable quantities as the
normalized transmitted intensity, $X=|x|^2$, and the normalized
incident intensity, $Y=|y|^2$. The transmitted and incident fields
are related by the state equation $y=x(1+2C)$. The atoms respond
to the external driving field by creating a polarization
$p=-<\sigma^+>$ that opposes that field and almost cancels it in
the low intensity limit. In terms of the normalized fields
(assuming equal phases as we are treating the resonant case), the
total intensity in the absence of atoms ($Y$) in terms of the
intensity in the presence of atoms ($X$) and the polarization is:
\begin{equation}
Y=|y|^2=|x+p|^2= |x|^2+2{\rm Re}(xp)+|p|^2= X+F \label{sse}.
\end{equation}
The total incoming energy $Y$ goes out as transmission $X$ or as
fluorescence $F=2{\rm Re}(xp)+|p|^2$. The fluorescence has two
components, one is the magnitude of the polarization and the other
a cross term between the intracavity field and the polarization
($2{\rm Re}(xp)$), similar to what we have in the equation of
motion of the atomic inversion (Eq.~\ref{inversion}). The
intensity escaping though the cavity mode has a contribution from
this term. It is rather difficult to separate from the drive and
from stimulated emission. However, it contains non-trivial
information about the atomic inversion, even in the case when
$<\sigma_z>=-1 + O(x^2)$, at the low intensity limit ($x<<1$).

\begin{figure}
\includegraphics{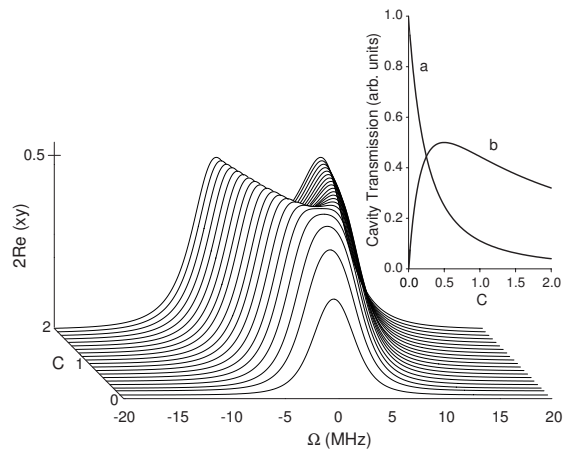}\caption{\label{figure1} Theoretical normalized
probe spectrum of spontaneous emission into the cavity mode
(equivalent to the atomic inversion spectrum) (2Re(xp)). for
$\kappa/2\pi=2.65$ MHz and $\gamma/2\pi=6$ MHz. Inset resonant
transmission for (a) the Vacuum Rabi spectrum and (b) atomic
inversion spectra.}
\end{figure}
\vspace*{-0.3cm}

Figure \ref{figure1} shows the theoretical calculation of the
transmitted fluorescence spectrum. This is the cross term between
the field and the polarization. It starts at zero for no atoms,
then grows to a maximum on resonance that then develops a doublet.
The peaks occur at $\Omega_{xp}=\pm
\sqrt{g^2N-(\kappa^2+(\gamma/2)^2)/2}$, a different value than
those for the transmitted spectrum, which splits with peaks at:
$\Omega_{X} = \pm
\sqrt{g^2N\sqrt{1+\left(\left(\gamma/g^2N\right)\left(\gamma/2+\kappa\right)\right)}-\left(\gamma/2\right)^2}$
\cite{gripp96a}.  As the cooperativity grows $C>>1$ the peaks of
the fluorescence spectrum into the cavity mode approach those of
the coupled atom-cavity system. The maximum on resonance
$\Omega=0$ occurs when a given drive $y$ excites to the upper
state an optimal number of atoms, before stimulated emission takes
over moving the spectrum away from the center into an
Autler-Townes-like doublet.

The inset in Fig.\ref{figure1} shows the normalized transmitted
spectra on resonance 
for the transmitted intensity and the spontaneous emission. The
transmitted intensity (a) starts at the peak of the transmission
of the empty cavity and decreases monotonically with $C$, while
the atomic inversion (b) starts at zero, grows and has a maximum
for $C=0.5$, and then decreases. The maximum coincides with the
place where the spectrum splits into two peaks. The two peaks
remain with a maximum value of 1/2 independent of C.

\begin{figure}[h]
\includegraphics{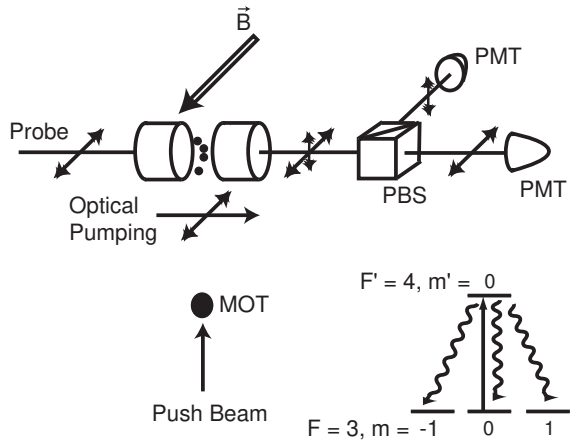}\caption{\label{figure2} Schematic of the experimental
apparatus. A polarizer at the output separates the two orthogonal
linear polarizations, one parallel to the driving field, the other
perpendicular and coming from the decay through $\Delta m=\pm 1$
spontaneous emission.}
\end{figure}

It is difficult to experimentally study the spontaneous emission
in cavity QED. Work in the past has focussed on geometries that
allow observation of the atoms from the side \cite{childs96}.
Another approach looks at the fluorescence into the mode of the
cavity with the atoms driven by a laser that propagates
perpendicular to the cavity axis \cite{zhu88,hennrich05}. We
follow Birnbaum \emph{et al.} \cite{birnbaum05} to directly access
a small part of the atomic inversion. We use the internal
structure of the atoms to inform us when a transmitted photon
originates in a fluorescence event (see Fig.~\ref{figure2}).
Instead of utilizing Rb atoms in their stretched states ($m_F=F$
with $\Delta m=1$) to form a closed two-level system when driven
with circularly polarized light, we prepare the atoms into the
$m_F=0$ ground state and drive the optical transition with $\pi$
polarization ($\Delta m=0$). We can then look at the light emitted
out of the cavity separating it into the two linear polarizations,
one parallel to the drive and the other orthogonal to the drive.
The presence of any light of orthogonal polarization signals that
it comes originally from a spontaneous emission event of an atom
that decays with $\Delta m \pm 1$.

\vspace*{-0.3cm}

The apparatus (see Fig. \ref{figure2}) consists of two main
components: The source of atoms and the cavity. Two lasers provide
the excitation radiation for the atomic source and for the cavity.
A titanium sapphire laser (Ti:Sapph) provides most of the light
needed for the experiment at 780 nm. The laser linewidth and
long-term lock are controlled using a Pound-Drever-Hall technique
on saturation spectroscopy of $^{85}$Rb. A second laser repumps
the atoms that fall out of the cycling transition in the trap.

A rubidium dispenser delivers Rb vapor to a magneto-optical trap
(MOT) in a glass cell 20 cm below a cubic chamber that houses the
cavity. The glass cell has a silane coating to decrease the
sticking of Rb to the walls and maximize the capture efficiency of
the MOT \cite{aubin03a}. We use a six beam configuration with 1/e
diameter of 20 mm (power) and 30 mW per beam.  A pair of
anti-Helmholtz coils generates a magnetic field gradient of 6 G/cm
and three sets of independent coils zero the magnetic field at the
trapping region.

The cavity defines a TEM$_{00}$ mode with two 7 mm diameter
mirrors with different transmission coefficients. The input
transmission (15~ppm) is smaller than the output (250~ppm) to
ensure that most of the signal escapes from the cavity on the
detector side. The cavity finesse for this arrangement is ${\cal
F}\approx 21\,000$ and its decay constant $\kappa/2\pi=2.6$ MHz.
The separation between the mirrors is 2 mm so the coupling
coefficient between the mode and the dipole transmission of Rb is
$g/2\pi=2$ MHz. The mirrors are glued directly to flat
piezo-electric-transducers (PZT) to control the length of the
cavity.

Our experimental system is in the intermediate regime of cavity
QED, where $g\approx ( \kappa , \gamma/2)$. Its rates  are $(g,
\kappa, \gamma/2)/2\pi=(2, 2.6, 3.0)$~MHz. Since single atom
cooperativity $C_1=0.25$ and the saturation photon number
$n_0=1.1$ the system requires about one photon in steady state to
become non-linear, but it starts to show the effects of
spontaneous emission at a much lower intensity.

We lock the cavity with a Pound-Drever-Hall technique using a 820
nm laser. This laser is locked to the stabilized 780 nm laser
using a transfer cavity. We separate the two wavelengths at the
output of the physics cavity with a grating, and use appropriate
interference filters to further ensure the separation of the two
colors.

We launch the atoms from the MOT towards the cavity with a
near-resonant probe beam from below. The repetition rate also sets
the number of atoms delivered to the cavity. We take data by
recording the transmitted light in the two orthogonal linear
polarizations for a fixed excitation frequency $\Omega$. There is
a slight non-degeneracy of the two orthogonal modes of less than
0.5 MHz (less than the full width at half maximum of the
transmission). The birefringence of the cavity is less than
$1\times 10^{-4}$ on its axis.

\begin{figure}[h]
\includegraphics{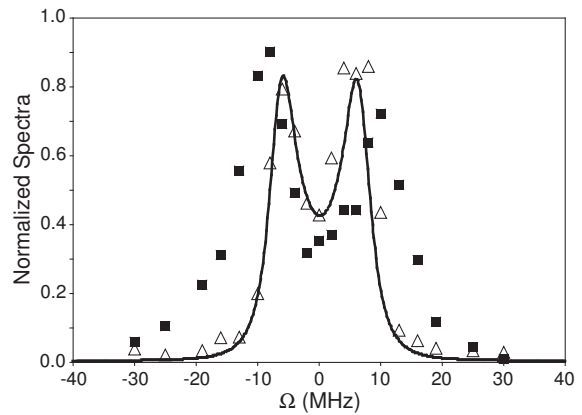}\caption{\label{figure3} Transmitted intensity
spectrum: vacuum Rabi in filled squares; atomic inversion in empty
triangles. The line shows the calculated spectrum for the
spontaneous emission into the mode of the cavity with the height
adjusted to match the normalized data.}
\end{figure}
\vspace*{-0.3cm}

The geometry that we use allows only $\pi$ transitions ($\Delta
m=0$) and no Faraday rotation of the light since an external
uniform magnetic field is aligned with the polarization direction
of the incoming light. The observed light at the orthogonal
polarization must come from spontaneous emission. This light is
emitted into the cavity mode so its detection is straightforward.
The input drive $Y$ is polarized horizontally to better than $1
\times 10^{-5}$ and aligned to the magnetic field to better than
$\pm 8$ degrees.

As each launch of atoms (every 150 ms) traverses the cavity we
record the transmission of both polarizations (parallel and
orthogonal) in a digital storage scope. We then change the
frequency of the driving laser and proceed to average over 200
launches of atoms. We extract from the raw data plots of the
transmission spectrum for a given $C$. Figure \ref{figure3} shows
the transmitted spectrum for the two orthogonal polarizations, the
one related to the intracavity field ($X$) in filled squares
(parallel polarization) and the other the one related to the
spontaneous emission in empty triangles (orthogonal polarization).
The peaks of the spectrum for $X$ are more separated than those of
the spontaneous emission. The calculated spectrum for the
spontaneous emission is normalized to match the data.

A detailed study of the susceptibility of this atomic system and
its response with two orthogonal polarizations is beyond the scope
of this letter and will be presented elsewhere
\cite{terraciano06}. We do not make any comments here about the
predictions of the shape of the spectrum based on the
simplifications of our model. Both polarizations have
contributions from spontaneous emission, but only in the
orthogonal polarization to the driving field the light coming from
the decay of an excited atom is clearly identified.
\begin{figure}[h] 
\includegraphics{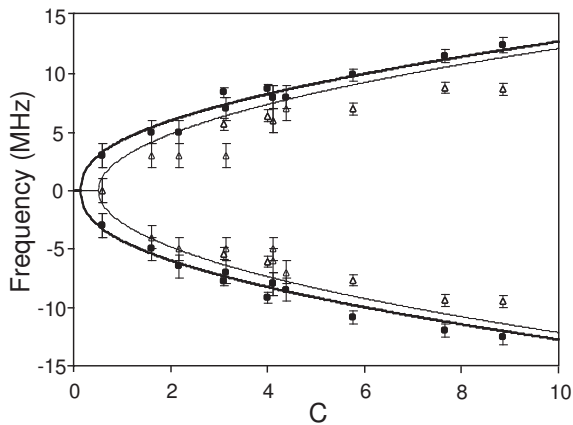}\caption{\label{figure4} Evolution of the position of the doublet
splitting in the transmitted intensity. Filled squares show the
intracavity intensity or vacuum Rabi splitting (parallel
polarization); empty triangles the spontaneous emission spectrum
(orthogonal polarization). The thick line is the prediction of
$\Omega_{X}$, while the thin line corresponds to $\Omega_{xp}$.}
\end{figure}
\vspace*{-0.3cm}

Figure~\ref{figure4} presents a comparison of the position of the
peaks in the measured spectra for both polarizations with the
predictions of our simple theory as function of $C$ which in our
case varies because of the change in the number of atoms. The
separation of the parallel polarization (cavity-atom doublet)
starts earlier that the one in the orthogonal polarization (atomic
inversion). For sufficiently high number of atoms, the positions
of the two peaks coalesce into the same doublet in our simplified
model. The overall horizontal scaling of this plot has been
adjusted based on the relationship between $C$ and $\Omega_{X}$,
and does not take into account any other broadening mechanisms.
Fitting $C$ to the size of the resonant transmitted light gives
consistent results within 30\% to those using $\Omega_{X}$.


The study of the spectrum of spontaneous emission into the mode of
a cavity QED system shows quantitatively different behavior from
the vacuum Rabi spectrum. The labelling of the photons by
polarization permits us to identify an emission out of the cavity
generated by an excited atom spontaneous decay. This will allow
measurements conditioned on the detection of a fluorescent photon
in the future. The specific quantum dynamics of this photon with
orthogonal polarization remain to be explored, in particular in
the regime where we can no longer neglect higher order
excitations.

This work was supported by NSF and NIST. We would like to thank H.
J. Kimble for his interest in this work. 

\end{document}